\documentclass{elsart}
\journal{New Astronomy}

\usepackage{natbib}

\usepackage{graphicx}
\usepackage{epsfig}

\usepackage{amssymb}

\def\wisk#1{\ifmmode{#1}\else{$#1$}\fi}

\def\deg    {\wisk{^\circ}}
\def\ddeg   {\wisk{{\rlap.}^\circ}}

\begin{document}

\begin{frontmatter}

%

\title{PAPPA: Primordial Anisotropy Polarization Pathfinder Array
\thanksref{SMD}}
\thanks[SMD]{This work was supported by the suborbital program
of the NASA Science Mission Directorate
under RTOP 188-02-16.}

\author[gsfc]{A. Kogut}
\author[gsfc]{D.T. Chuss}
\author[gsfc,ssai]{D. Fixsen}
\author[gsfc]{G.F. Hinshaw}
\author[gsfc,ssai]{M. Limon}
\author[gsfc]{S.H. Moseley}
\author[gsfc,ssai]{N. Phillips}
\author[gsfc,gst]{E. Sharp}
\author[gsfc]{E.J. Wollack}
\author[gsfc]{K. U-Yen}
\author[gsfc,mei]{N. Cao}
\author[gsfc]{T. Stevenson}
\author[gsfc]{W. Hsieh}
\author[upenn]{M. Devlin}
\author[upenn]{S. Dicker}
\author[upenn]{C. Semisch}
\author[nist]{K. Irwin}
\address[gsfc]{NASA Goddard Space Flight Center, Greenbelt, MD 20771}
\address[ssai]{Science Systems and Applications, Inc.}
\address[gst]{Global Science and Technology}
\address[mei]{MEI Technologies}
\address[upenn]{University of Pennsylvania, Philadelphia PA}
\address[nist]{National Institute of Standards and Technology, 
Boulder CO}

\begin{abstract}
The Primordial Anisotropy Polarization Pathfinder Array (PAPPA)
is a balloon-based instrument to measure the polarization
of the cosmic microwave background
and search for the signal from gravity waves
excited during an inflationary epoch
in the early universe.
PAPPA will survey a 
$20\deg ~\times ~20\deg$ patch
at the North Celestial Pole
using 32 pixels 
in 3 passbands centered at 89, 212, and 302 GHz.
Each pixel uses MEMS switches 
in a superconducting microstrip transmission line
to combine
the phase modulation techniques
used in radio astronomy
with the sensitivity of transition-edge superconducting bolometers.
Each switched circuit modulates the incident polarization
on a single detector,
allowing nearly instantaneous characterization 
of the Stokes I, Q, and U parameters.
We describe the instrument design and status.
\end{abstract}

\begin{keyword}
cosmology 
\sep cosmic microwave background 
\sep polarization 
\sep instrumentation
\end{keyword}

\end{frontmatter}

\section{Introduction}
Linear polarization of the cosmic microwave background (CMB)
carries the oldest information in the universe.
Quantum fluctuations of the space-time metric (gravity waves)
excited only $10^{-35}$ seconds after the Big Bang
impart a distinctive signature to the CMB polarization.
Detecting this signal
probes physics at energies near 
Grand Unification
and could provide insight on
quantum gravity.

\begin{figure}[b]
\centerline{
\psfig{file=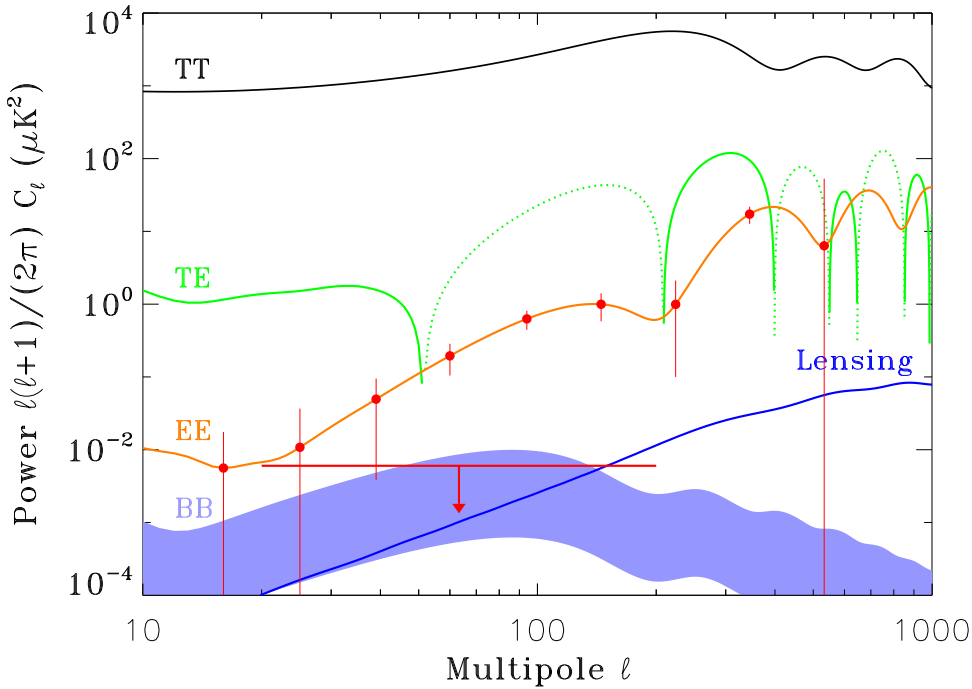,width=4.0in,angle=90}}
\caption{
Angular power spectra of the temperature and polarization anisotropy.
The blue band shows the gravity-wave signal
at tensor/scalar ratio $0.01 < r < 0.1$.
The red line shows the PAPPA B-mode sensitivity after projecting out
foregrounds.
PAPPA will provide a precise measurement of the E-mode polarization
and can detect B-mode polarization at levels r = 0.1.
}
\label{power_spectra_fig}
\end{figure}

CMB polarization results from Thomson scattering of CMB photons
by free electrons.
Scattering of an isotropic radiation field 
produces no net polarization,
but a quadrupole moment in the incident radiation yields
a polarized signal.
The required quadrupole can result from
either intrinsic fluctuations in the radiation field itself,
or the differential redshift as gravity waves
propagate through an isotropic medium.
Temperature or density perturbations are scalar quantities;
their polarization signal must therefore be curl-free.
Gravity waves, however, are tensor perturbations
whose polarization includes both gradient and curl components.
In analogy to electromagnetism,
the scalar and curl components are
often called ``E'' and ``B'' modes.
Only gravity waves induce a curl component:
detection of a B-mode signal in the CMB polarization field
is recognized as a ``smoking gun'' signature of inflation,
testing physics at energies inaccessible through any other means
\cite{kamionkowski/etal:1997,seljak/zaldarriaga:1997}.

Figure \ref{power_spectra_fig} shows the power spectra
for the temperature and polarization.
E-mode polarization depends on the quadrupole anisotropy
of the temperature distribution
seen by each scatterer within its horizon.
The temperature distribution is observed to be anisotropic
and the physics of Thomson scattering is well understood;
hence,
the CMB {\it must} be partially polarized
with E-mode amplitude predictable from the temperature anisotropy
and background cosmology.
Polarization at the predicted amplitude
was first observed by the DASI collaboration
\cite{kovac/etal:2002}
and later confirmed by multiple experiments.

Recent results from the WMAP mission
suggest that an observable gravity-wave signature should exist.
Quantum-mechanical fluctuations in the inflaton field
produce both density perturbations and gravity waves.
The transition out of the inflationary epoch
breaks pure scale invariance;
the resulting signature in the scalar density perturbations
can be used to estimate the accompanying signal
from tensor gravity waves.
Given the WMAP measurement of a deviation from scale invariance,
the simplest inflation models predict
gravity waves
with tensor/scalar ratio in the range $0.01 < r < 0.16$
\cite{spergel/etal:2006,
linde:2005,
lyth/riotto:1999}.
The corresponding B-mode signal
has RMS amplitude 30--100 nK 
at degree angular scales.
Signals at this level can be detected by a dedicated polarimeter,
providing a crucial test of inflationary physics.

\begin{figure}[b]
\centerline{
\psfig{file=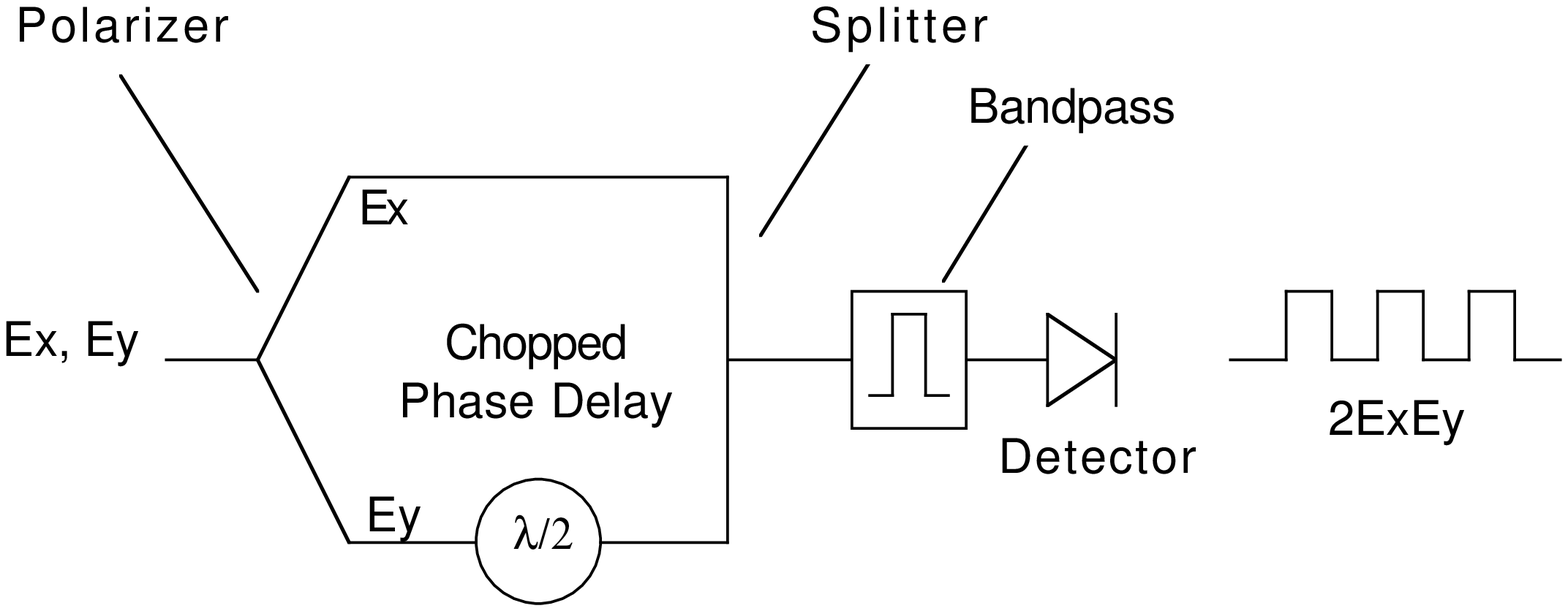,width=4.0in}}
\caption{Schematic planar polarimeter.
A phase switch 
introduces a half-wave delay
in one arm
before the signals
are combined and detected.
The detector output
has a dc term proportional to the unpolarized intensity (Stokes I)
plus an ac term at the switch frequency
proportional to the linear polarization.
}
\label{iu_schematic_fig}
\end{figure}

Detecting the gravity-wave signature
in polarization will be difficult.
The first challenge is sensitivity.
The gravity-wave signal is faint
compared to the fundamental sensitivity limit
imposed by photon arrival statistics.
Even noiseless detectors suffer from this photon-counting limit;
the only solution is to use multiple independent detectors.
Equally important is the control of systematic errors.
Aliasing of power from either unpolarized sources
or the dominant E-mode polarization
into a spurious B-mode pattern
could overwhelm the primordial signal.
The Primordial Anisotropy Polarization Pathfinder Array (PAPPA)
uses phase-sensitive switching techniques
to combine
the sensitivity of transition-edge superconducting detectors
with
rigorous control of potential systematic errors.
With no macroscopic moving parts,
PAPPA is scalable to the kilo-pixel arrays
anticipated for an eventual space mission
dedicated to CMB polarimetry.

\section{Instrument Description}
PAPPA will consist of an array of 32 ``polarimeters-on-a-chip''
in 3 frequency bands centered at 89, 212, and 302 GHz.
Each pixel uses phase-delay microcircuits
coupled to the sky using
corrugated feed horns
and a 60-cm off-axis primary mirror
to produce 0\ddeg5~beams on the sky.
Each circuit periodically injects a half-wave phase delay
between orthogonal polarization components
launched into superconducting microstrip,
which are then combined and detected
using transition-edge superconducting bolometers.
A planar polarimeter allows rigorous control of systematic errors.
The design is easily implemented in microstrip,
so that all elements (including the phase switch)
are at the same temperature as the detector.
The modulation can be very fast (100 Hz),
reducing the effects of slow drifts in the readout electronics.
The synchronously demodulated output is proportional
only to the linearly polarized part of the sky signal --
uncertainties in the detector responsivity
only affect the estimated amplitude of any polarized signal,
but can {\it not} generate a false polarization signal
from an unpolarized sky.
The entire polarimeter can be fabricated 
using photolithographic techniques,
making it well suited for large array formats.

\subsection{Polarization Modulation}
Consider an incident plane wave
\begin{equation}
E = E_x \cos(kz - \omega t - \phi_x)
  + E_y \cos(kz - \omega t - \phi_y)
\label{plane_wave_eq}
\end{equation}
with component amplitude
$E_x$ and $E_y$ in the $\hat x$ and $\hat y$ directions, respectively.
We can describe the polarization state using the Stokes parameters
\begin{eqnarray}
{\rm I} =  \langle E_x^2 + E_y^2 \rangle & &
{\rm U} = 2 \langle E_x E_y \cos(\phi_x - \phi_y) \rangle \nonumber \\
{\rm Q} = \langle E_x^2 - E_y^2 \rangle  & &
{\rm V} = 2 \langle E_x E_y \sin(\phi_x - \phi_y) \rangle 
\label{stokes_def_eq}
\end{eqnarray}
where the brackets indicate integration over a time
long compared to the frequency $\omega$.
If the phases $\phi_x$ and $\phi_y$ 
differ by a multiple of $\pi$,
then the direction of the $E$ vector in the xy plane is fixed in time,
and the wave is linearly polarized.

\begin{figure}[t]
\centerline{
\psfig{file=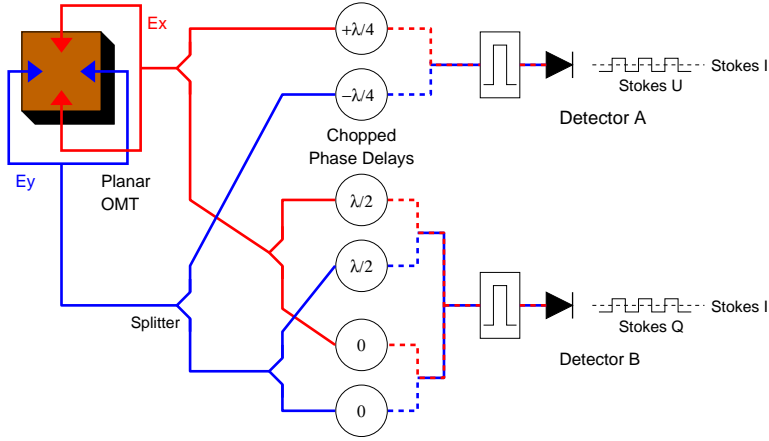,width=4.0in}}
\caption{Symmetrized microstrip polarimeter for PAPPA.}
\label{iqu_schematic_fig}
\end{figure}

Reliable characterization
of faint polarization
requires modulating the signal on a single detector
on time scales short compared to instrumental drifts, $1/f$ noise,
or the motion of the beams across the sky.
A phase-sensitive design
allows measurement of linear polarization
while maintaining rigorous control of systematic errors.
Figure \ref{iu_schematic_fig}
illustrates the concept.
A polarizing microstrip element
(e.g., orthogonal probes in a resonant structure)
launches voltages into two arms of the polarimeter,
proportional to the electric field amplitudes
$E_x \cos(kz - \omega t - \phi_x)$
and 
$E_y \cos(kz - \omega t - \phi_y)$.
A half-wave phase switch 
alternately injects a phase delay $0$ or $\pi$
in one arm.
The voltages from the two arms are then combined
before square-law detection.
When the phase switch is off, the voltages in the two arms are
\begin{eqnarray}
V_x & = & E \cos\alpha \cos(\omega t - \phi_x) \nonumber \\
V_y & = & E \sin\alpha \cos(\omega t - \phi_y) 
\label{iu_v_off}
\end{eqnarray}
where
$E = \sqrt( E_x^2 + E_y^2 )$
and
$\alpha$ is the angle between the linearly polarized incident field
and the $xy$ coordinate system of the polarimeter.
The detector power is
\begin{eqnarray}
P_{\rm off} & = & \langle (V_x + V_y)^2 \rangle \nonumber \\
 & = & E^2( 1 + 2\cos\alpha\sin\alpha )
\label{P_off}
\end{eqnarray}
up to an uninteresting constant phase.
When the phase switch is on,
it introduces an additional half-wave path length so that
$V_y = E \sin\alpha \cos(\omega t - \phi_y + \pi)
= -E \sin\alpha \cos(\omega t - \phi_y)$.
The detector power is then
\begin{eqnarray}
P_{\rm on} & = & E^2( 1 - 2\cos\alpha\sin\alpha )
\label{P_on}
\end{eqnarray}
As the switch chops, the detector produces a slowly-varying offset
\begin{eqnarray}
P_{dc}  & = & (P_{\rm on} + P_{\rm off}) / 2 \nonumber \\
	& = & E^2 ~~( {\it i.e.}~{\rm Stokes ~I} ) 
\label{dc_power}
\end{eqnarray}
proportional to the unpolarized intensity,
plus a rapidly modulated term
\begin{eqnarray}
P_{ac}  & = & (P_{\rm on} - P_{\rm off}) / 2 \nonumber \\
	& = & 2 E^2 \cos\alpha \sin\alpha \nonumber \\
        & = & 2 E_x E_y ~~( {\it i.e.}~{\rm Stokes ~U} ) 
\label{ac_power}
\end{eqnarray}
proportional to the linear polarization (Stokes U).
We thus unambiguously separate the 
polarized and unpolarized components
in a single measurement with a single detector.

\begin{figure}[b]
\centerline{
\psfig{file=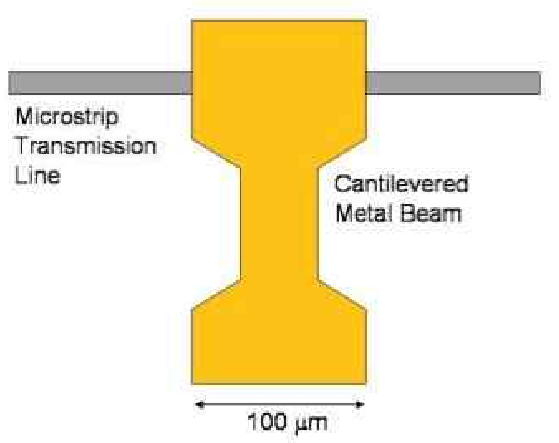,height=2.2in}
\psfig{file=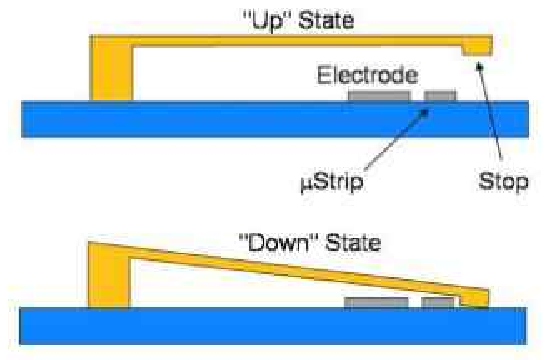,height=2.2in}}
\caption{Schematic showing MEMS phase switch.
(Left) Top view showing metal beam cantilevered above transmission line.
(Center) Side view showing the beam
in the ``up'' and ``down'' positions.
Metal posts maintain a 0.25 $\mu$m gap in the ``down'' position
to prevent physical contact between the beam and transmission line.
}
\label{mems_schematic}
\end{figure}

The concept can readily be extended to provide
simultaneous measurements of the Stokes I, Q, and U parameters.
Figure \ref{iqu_schematic_fig} shows 
the circuit implemented for PAPPA.
A polarizing element again launches voltages onto microstrip 
components proportional to the $x$ and $y$ components of the 
incident electric field.
Each component is then split to separate detection chains.
Detector A (top) is identical to the IU polarimeter 
in Figure \ref{iu_schematic_fig}.
Detector B has an additional pair of phase switches
run out of phase with each other
(e.g., S1 has phase delay 0 when S2 has delay $\lambda/2$,
or else
S1 has phase delay $\lambda/2$ when S2 has delay 0).
As switches S1 and S2 chop,
detector B has a dc term proportional to Stokes I,
and a modulated term 
$P_{ac} = E_x^2 - E_y^2$
proportional to the Stokes Q parameter.
By simultaneously measuring I, Q, and U,
PAPPA fully characterizes the incident linear polarization.
This decoupling of the polarization measurement
from the instrument orientation or beam scan motion
greatly reduces the problem
of aliasing unpolarized anisotropy into 
a spurious polarization signal.

Microstrip delay lines of different path length
provide a simple, convenient way to produce such a phase shift.
Switched capacitors couple to the transmission line
to short out a delay line of length $\lambda$/2
or $\lambda/4$.
Micro Electro Mechanical Systems (MEMS) switches
are a demonstrated technology
capable of producing the required modulation.
MEMS switches are miniature surface micromachined components
providing controlled motion over a short distance
to create either
an open circuit or a short across a transmission line.
Figure \ref{mems_schematic}
shows a MEMS capacitive switch
for the sub-mm  phase modulator.
It consists of a metal beam
300 $\mu$m $\times$ 100 $\mu$m $\times$ 1 $\mu$m thick
cantilevered 1 $\mu$m above the transmission line.
An electrostatic voltage applied to a pull-down electrode
pulls the beam down
until it is stopped by metal posts
some 0.25 $\mu$m above the transmission line.
The change in gap height between the ``up'' and ``down'' states
determines the change in capacitance;
we achieve acceptable performance with ratio 4:1 or greater.
The design is tolerant to changes in the ``up'' state capacitance.

\begin{figure}[b]
\centerline{
\psfig{file=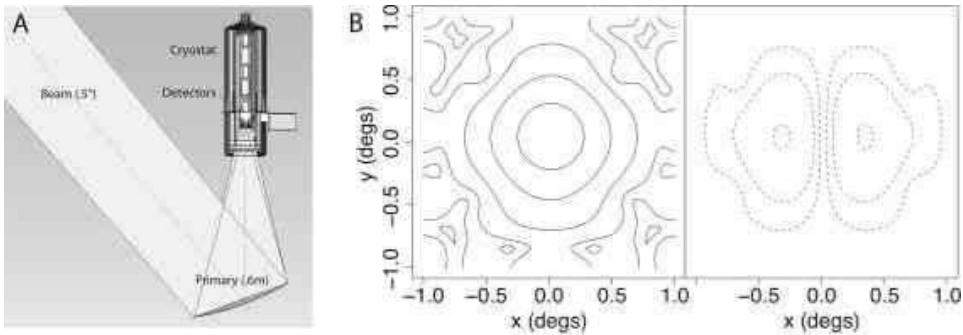,width=5.0in}
}
\caption{(Left) 
Optical design showing the 60 cm off-axis primary
with the cryostat at prime focus.
(Right)
Calculated co-polar and cross-polar beam patterns
for a pixel at the edge of the 89 GHz array.
Co-polar contours are -3, -10, -20, -40, and -50 dB;
cross-polar contours are -30, -40, and -50 dB.
}
\label{optics_fig}
\end{figure}

\subsection{Optics and Cryostat}
Figure \ref{optics_fig}
shows the PAPPA optical design.
PAPPA uses a single 60 cm off-axis primary mirror
to focus light from the sky onto the cryogenic focal plane.
An array of 1.5K corrugated feed horns
at the primary focus
couples light from the telescope to the 100 mK microstrip.
Use of a single warm reflector
minimizes thermal loading on the detectors,
providing a significant reduction in noise.
The illumination of the feedhorns on the primary mirror
allows -30 dB edge taper.
Figure \ref{optics_fig} shows the calculated 
co-polar and cross-polar
response of the beams on the sky.
The cross-polar response is small:
end-to-end simulations show that
spurious B-modes 
resulting from power aliased out of the dominant E-mode signal
by beam imperfections
will be more than 2 orders of magnitude
below the PAPPA noise limit.

\begin{figure}[b]
\centerline{
\psfig{file=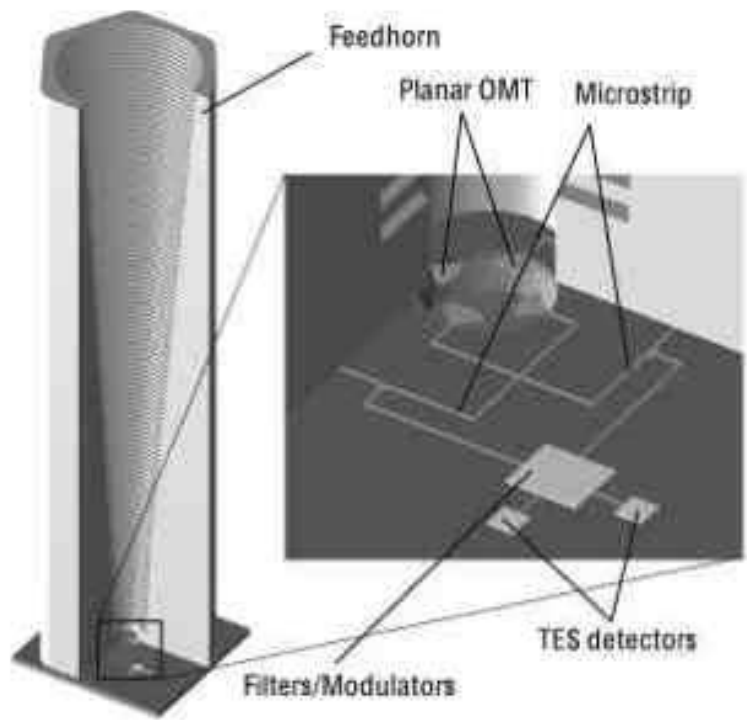, width=4.0in}
}
\caption{
Schematic diagram of a single PAPPA pixel
showing the feedhorn,
planar OMT,
and superconducting microstrip.}
\label{pixel_fig}
\end{figure}

\begin{figure}[b]
\centerline{
\psfig{file=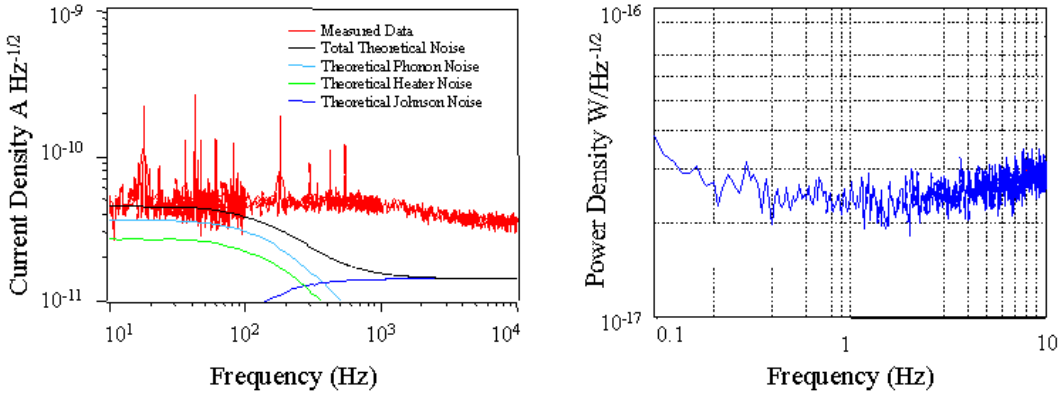,width=5.8in}}
\caption{
Measured noise performance in GSFC/NIST TES bolometer.
(Left) Measured and theoretical current noise.
No free parameters are used to generate theoretical predictions.
Structure in the measured noise below 1 kHz is due to line pickup
in the measurement system.
The measured noise is within a few percent of the thermodynamic limit
in the signal bandpass (below 100 Hz).
(Right) Low-frequency power spectral density.
The noise is white with 1/f knee 0.1 Hz.
Lower operating temperature 0.1 K
will bring the NEP below $10^{-17}$ W Hz$^{-1/2}$.
}
\label{tes_noise_fig}
\end{figure}

Figure \ref{pixel_fig}
shows a single PAPPA pixel.
A planar ortho-mode transducer (OMT)
in a resonant cavity behind the feedhorn array
launches orthogonal polarizations 
into superconducting niobium microstrip lines.
A quarter-wave backshort 
reflects incident radiation
onto orthogonal microstrip probes
while simultaneously serving as a thermal gap
between the 1.5 K feedhorn array
and the 100 mK microstrip circuit.
A gap also avoids impulsive heating of the focal plane and detectors
from cosmic ray hits within the feedhorn volume.
We reconcile the wavelength-dependent spacing for the backshort
with the use of 3 different frequency bands
by breaking the focal plane assembly
into 3 separate sub-assemblies,
one for each band.
Each polarimeter will use
transition-edge superconducting (TES) bolometers 
at 100 mK
as square-law detectors for the phase circuit.
Figure \ref{tes_noise_fig} shows
the measured noise from a prototype Mo:Cu device.
We measure noise within a few percent of the thermodynamic limit,
with noise equivalent power 
${\rm NEP} = 2.3 \times 10^{-17}$ W Hz$^{-1/2}$
and time constant 
$\tau \sim ~1$ ms.

\begin{figure}[b]
\centerline{
\psfig{file=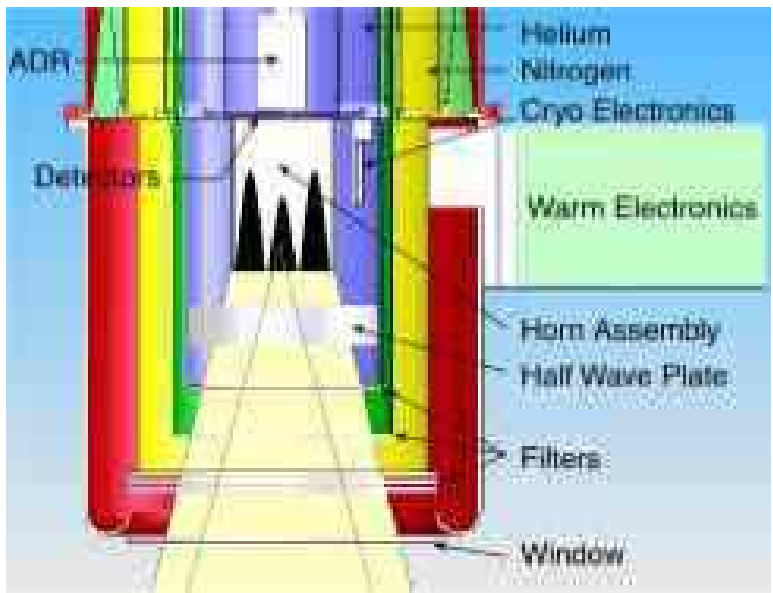,width=4.0in}}
\caption{Schematic of the cryostat
showing the cold optics and detector location.
}
\label{cryostat_fig}
\end{figure}

\begin{figure}[b]
\centerline{
\psfig{file=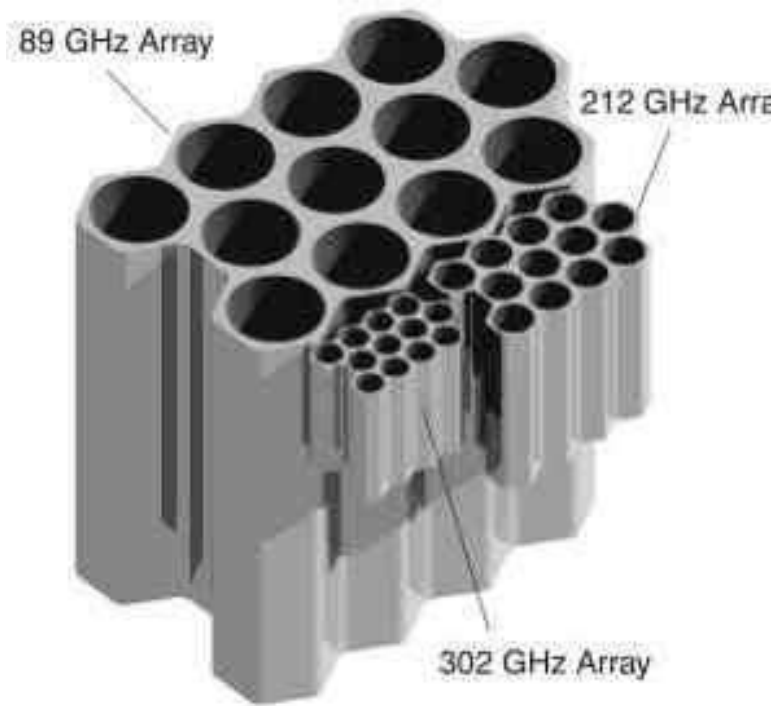,width=4.0in}}
\caption{Feedhorn arrays couple light from the telescope
to the polarimeter microcircuit.
Each array is monochromatic.
}
\label{feedhorn_fig}
\end{figure}

Figure \ref{cryostat_fig}
shows the layout of the cold focal plane.
The 10 cm diameter focal plane
lies at the prime focus of the off-axis primary mirror
and is maintained at 100 mK
using a 3-stage ADR 
sunk to a liquid helium reservoir at 1.5 K.
The ADR provides 21 $\mu$W cooling power at 100 mK.
Blocking filters limit
the infrared heat load
from the 20 cm window to the focal plane.
Three separate monochromatic feedhorn arrays 
(Figure \ref{feedhorn_fig})
built using platelet techniques
couple light onto the polarimeter microcircuit.

\subsection{Sensitivity}
Table \ref{system_table}
summarizes the PAPPA instrument sensitivity.
It consists of an array of 32 bolometers
in three frequency bands
centered at 89, 212, and 302 GHz.
Each bolometer detects the signal
from a horn-coupled planar microstrip polarimeter.
Coupling to the microstrip components
is efficient ($\epsilon \sim 0.95$);
the total efficiency
is dominated by the insertion loss of the microstrip bandpass filters.
We achieve NEP $2 \times 10^{-17}$ W Hz$^{-1/2}$
in existing bolometers;
operating at lower temperature
and with better thermal isolation in the support legs
will reduce the NEP below $10^{-17}$ W Hz$^{-1/2}$.
PAPPA uses a single ambient-temperature mirror
so that the power incident on each bolometer
is dominated by emission from the CMB:
PAPPA operates within 70\% of the CMB photon noise limit.
The noise equivalent temperature of a single bolometer
observing a polarized CMB source
is 26 and 32 $\mu$K Hz$^{-1/2}$ for the 89 GHz and 212 GHz channels,
and 236 $\mu$K Hz$^{-1/2}$ for the 302 GHz channel
where the CMB is weaker and the dust foreground is brighter.
The number of bolometers in each frequency band
is chosen to provide nearly equal signal-to-noise ratio
in each band
to optimize discrimination
between CMB and galactic foregrounds.

\begin{table}[t]
{
\small
\label{system_table}
\begin{center}
\begin{tabular}{l c c c}
\multicolumn{4}{c}{Table 1: Instrument Sensitivity} \\
\hline 
Parameter			& 	Band 1	&	Band 2	& Band 3 \\
\hline 
Center Frequency (GHz)		& 	89	&	212	&  302 	\\
Band Width ($\Delta\nu / \nu$)	& 	0.44	&	0.37	& 0.10 	\\
Optical Efficiency		&	0.7	&	0.7	& 0.5	\\
Bolometer NEP (W Hz$^{-1/2}$)	&  	$1.0 \times 10^{-17}$ 
				&	$1.0 \times 10^{-17}$
				&	$1.0 \times 10^{-17}$	\\
Total NEP (W Hz$^{-1/2}$)	&  	$1.3 \times 10^{-17}$ 
				&	$1.3 \times 10^{-17}$
				&	$1.2 \times 10^{-17}$	\\
Temperature NET ($\mu$K Hz$^{-1/2}$) &	18	&	23	&	167 \\
Polarimeter NEQ ($\mu$K Hz$^{-1/2}$) &	26	&	32	&	236 \\
Number of Detectors		& 	12	&	12	& 	8   \\
Band NEQ ($\mu$K Hz$^{-1/2}$)	&	7	&	9	&	83   \\
\hline 
\end{tabular}
\end{center}
}
\end{table}

PAPPA will scan a $20\deg \times 20\deg$ patch of sky,
using a raster strategy along lines of constant elevation.
A deep scan on a limited patch of sky
provides greatest sensitivity
to the B-mode signal from gravity waves
\cite{lewis/etal:2002}.
Sky rotation as the gondola tracks the patch
provides cross-linking of pixels in multiple directions.
Twenty hours of data from a turn-around North American balloon flight
provide sensitivity of 1.3 $\mu$K
per 0\ddeg5 ~beam.
Limited sky coverage
creates ambiguities 
when decomposing
local Stokes parameters Q and U
into the global E and B parameters.
We simulate this effect over the PAPPA sky coverage
and include it in all sensitivity estimates.
PAPPA
will characterize the E-mode power spectrum
for multipoles $l < 400$
and detect or limit B-modes
corresponding to tensor/scalar ratio $r = 0.1$.
The PAPPA sensitivity 
is more than a factor of 5 better than existing upper limits
and begins to probe 
the inflationary signal levels
predicted by the WMAP detection of deviations
from pure scale invariance.

\section{Status and Plans}
A prototype of the PAPPA payload will fly in September 2007.
This initial payload 
will consist of a horn-coupled microstrip array
using a rotating half-wave plate
instead of phase switches
to provide the necessary polarization modulation 
on each detector.
A later flight will incorporate the full micro-circuit
including the MEMS phase switches
to demonstrate the full array.

\begin{thebibliography}{99}
\frenchspacing

\bibitem{kamionkowski/etal:1997}
Kamionkowski, M., Kosowsky, A., and Stebbins, A., PRD, {\bf 55}, 7368 (1997).

\bibitem{kovac/etal:2002}
Kovac, J.\ M., et al., Nature, {\bf 420}, 772 (2002).

\bibitem{lewis/etal:2002}
Lewis, A., Challinor, A., and Turok, N., PRD, {\bf D65} 023505 (2002).

\bibitem{linde:2005}
Linde, A.\ D., New Astronomy Review, {\bf 49}, 35 (2005).

\bibitem{lyth/riotto:1999}
Lyth, D.\ H., and Riotto, A.\, Phys. Rep., {\bf 314}, 1 (1999).

\bibitem{seljak/zaldarriaga:1997}
Seljak, U., and Zaldarriaga, M., PRL, {\bf 78}, 2054 (1997).

\bibitem{spergel/etal:2006}
Spergel, D.~N., et al., ApJ, submitted
(preprint astro-ph/0603449).

\end{thebibliography}
\end{document}